# Teaching methods are erroneous: approaches which lead to erroneous end-user computing

Mária Csernoch, Piroska Biró
4028 Kassai út. 26. Debrecen, Hungary
csernoch.maria@inf.unideb.hu, biro.piroska@inf.unideb.hu

**ABSTRACT**

*If spreadsheets are not erroneous then who, or what, is? Research has found that end-users are. If end-users are erroneous then why they are? Research has found that responsibility lies with human beings' fast and slow thinking modes and the inappropriate way they use them. If we are aware of this peculiarity of human thinking, then why do we not teach students how to train their brains? This is the main problem, this is the weakest link in the process; teaching. We have to make teachers realize that end-users are erroneous because of the erroneous teaching approaches to end-user computing. The proportion of fast and slow thinking modes is not constant, and teachers are mistaken when they apply the same proportion in both the teaching and end-user roles. Teachers should believe in the incremental nature of science and have high self-efficacy to make students understand and appreciate science. This is not currently the case in ICT and CS, and it is high time fundamental changes were introduced.*

## 1 INTRODUCTION

Research focusing on spreadsheet analysis has come to the conclusion that almost without exception spreadsheet documents, – more than 90% of them – have various types of errors, and that these errors – along with the extremely high level of human and computer resources required to administer the documents (EuSpRIG, Panko, 2008; Powel et al., 2008; Thorne, 2010) – cause serious financial losses (Van Deursen & Van Dijk, 2012). Speaking generally, it has been accepted that spreadsheets are erroneous. However, Panko (2013) claimed that it is not spreadsheets which are erroneous but rather the end-users who create the documents. He explains that one of the reasons for making mistakes is the overuse of attention mode (ATM) thinking compared to automatic mode (AUM) thinking (Panko, 2013, 2015; Maynes, 2015; Kahnemann, 2011). This finding is closely related to Kelemen's, who claims that there is unreliability in metacognitive accuracy, while both memory and confidence are usually consistent between tasks (Kelemen et al., 2000).

Compared to Panko, our research group took several further steps by analyzing the different metacognitive computer problem solving approaches, the problem solving approaches of end-users, the mathability level of software tools, cognitive load theory, the teaching methods applied in end-user teaching and training, the textbooks and coursebooks, teacher education, as well as several informatics and computer science curricula. Considering all these different approaches and their connection to spreadsheets, we have found that one of the main reasons spreadsheet users make mistakes is that teaching methods and materials are erroneous. Consequently, until we transform end-user-teaching approaches, nothing will change. In the present paper we focus on the educational aspect of the TEAM (Tools Education Audit Management) Approach (Chadwick, 2002). We argue that we have both the theoretical background and the teaching tools needed to introduce concept- and algorithmic-based spreadsheet management as an effective tool in end-user computing.

## 2 PROBLEMS TO DEAL WITH

We are faced with a high number of problems in end-user-teaching. (1) As mentioned in the Introduction, Panko (2013), based on his research on cognitive science, claimed that most spreadsheet errors are due to ATM thinking. Consequently, we have to develop end-users' AUM thinking to reduce spreadsheet errors. (2) Panko & Port (2013) have also claimed that "[end-user computing] … seems to be invisible to the central corporate IT group, to general corporate management, and to information systems (IS) researchers." (Panko & Port, 2013; Burnett, 2009). (3) "The public image of computer science does not reflect its true nature. The general public and especially high school students identify computer science with a computer driving license. They think that studying computer science is not a challenge, and that anybody can learn it. Computer science is not considered a scientific discipline but a collection of computer skills." (Hromkovic, 2009). These misleading opinions are openly expressed by Gove (2012) and Bell & Newton (2013). "…children bored out of their minds being taught how to use Word and Excel by bored teachers…" (Gove, 2012). "…a collection of low-level routine knowledge such as how to format pages in a word processor, or how to make tables in HTML." (Bell & Newton, 2013).

The following three problems are straightforward consequences of problem (3) mentioned above. (4) Teaching materials – textbooks, coursebooks, recently published e-materials, etc. – do not support the development of computational thinking, which Wing claimed was the newly emerged basic skill of the digital era. "Computational thinking is a fundamental skill for everyone, not just for computer scientists. To reading, writing, and arithmetic, we should add computational thinking to every child's analytical ability." (Wing, 2006). These teaching materials are mostly out of context, beyond this they focus on the details of the tools, are written in cookbook style (Appendix 4) – giving sequences of clicks as instructions (Angeli, 2013) or, in reference style, – replicate reference materials. (5) Teachers, almost unconditionally, accept these teaching materials, and the approaches and methods which they suggest. (6) Teacher education is not prepared for the challenges of the digital era, neither in the case of teachers of informatics and computer sciences, nor non-professionals (Csernoch, 2015; European Schoolnet, 2011, 2013, 2015).

## 3 TEACHING MATERIALS

By analyzing numerous spreadsheet teaching materials we have found that these books, on-line courses, and the teachers who follow them are one of the main reasons for failure. We focused on general purpose informatics coursebooks with a section on spreadsheets, as well as on books specializing in spreadsheets (Csernoch et al., 2014).

The analyses of these books revealed that they do not fulfill the requirements of the general concepts and basic rules of informatics textbooks detailed in the paper by Freiermuth et al. (2008), but rather follow the "…the misleading concepts of computer science education that were broadcasted in many countries as the consequences of the emphasis created by the fast development of information technologies."

In general coursebooks the spreadsheet sections/chapters focus on formatting details: how to color cells and borders, how to change fonts and font styles, etc., creating diagrams, and providing different lists of functions. What we miss in these books is real world problems and problem solving. The tasks, if there are any, are only fabricated examples based on non-existing or fake tables (Appendix 3), focusing on the details of the language, mostly listing functions and their arguments (Csernoch et al., 2014; Appendix 5: students' and books' list of functions, S and B, respectively). The only exception found with real-world problems is Gross et al.'s book (2014), however they also present a fictional company and detail general informatics. None of the books corrects the errors contained in the wizards, helps, and/or references, as, for example, is shown by the arguments of the MATCH() and the IF() functions. The MATCH() function (MATCH function, nd) does not accept any array as an argument, and only a one-dimensional array (vector) serves as the lookup array. In the reference of the IF() function (IF function, nd) the 'logical test', 'logical expression', 'condition' is named as the first argument, but untrained end-users do not understand these expressions, while they are familiar with the notion of 'yes/no question'. In a similar way, end-users do not understand the match-type argument of the MATCH() function, since the reference is based on the different selection algorithms (MATCH function, nd); however, they understand the concepts 'descending', 'ascending', and 'no order', which are necessary to select the correct match-type.

The spreadsheet books we analyzed turned out to be ill-named spreadsheet books. Instead of teaching spreadsheets they teach general ICT skills. In addition, the same introductory chapters on managing text, presentations, and spreadsheets, etc. can be found in all the birotical – office applications – coursebooks (Appendices 1, 2, and 4).

In these spreadsheet books a great range of basic knowledge about informatics is detailed at great length, just like in general coursebooks (Apppendix 4. We have to mention here, that copying is four-step process, which last two are merged in the example). Beyond these, the newest features are emphasized and only extremely short sections deal with functions. The only exceptions were the books by Walkenbach (2002, 2010) and Advanced Excel Essentials (Goldmeier, 2014). Walkenbach works with formulas, functions, and even with array formulas. He mentions that one of the advantages of array formulas over copying formulas is that they reduce the vulnerability of spreadsheets. Goldmeier's book is much less conceptualized, seeming to feature ideas which pop up randomly, without a clear understanding of the concepts of problem solving and arrays.

Considering the contents analyzed, we have come to the conclusion that these books represent a paradox. First of all, their content and style have not changed over the last two decades. This means that even the newest books published in the digital era, in which the greatest number of generations of end-users are using computers, explain how to start a program, and how to open and save files (Appendices 1 and 2). This information is either completely unnecessary or presupposes that the readers' digital competence is at an extremely low level. The paradox is here: if end-users are able to handle files, these contents should not be in spreadsheet books; if end-users do not know how to handle files, they should be taught, but this is not the task of spreadsheet textbooks. Anyway we claim that these chapters should be omitted from spreadsheet textbooks. In a similar way, formatting and typographic details, along with knowledge about styles, do not constitute spreadsheet knowledge. These elements should also be part of end-users' digital competence. In general, these books mainly focus on the interface and mix basic ICT knowledge with the most recent features of the programs, and miss the essence of spreadsheets.

On the other hand, at the very beginning of these books readers are overwhelmed with subjects for which they are not prepared. Is it worth explaining all the different data types for beginners in one group? No, it is just waste of time and energy. Instead, we can open files with obvious examples, students can analyze them and recognize the different data types – and not all of them at the same time. Or we can open a file with mismatched data types and students will see it even more clearly. Explaining references for beginners? They will not understand them. No macros in introductory courses, please!

What is not included in these books? There is no real problem solving in coursebooks, not even in online tutorials (e.g. for a sample of formulas on an empty table see Appendix 3), where adding tables with authentic content would not be a problem. These books are weak copies of references. The books do not mention how to design content (Angeli, 2013), how to solve problems, or how to build algorithms (Hubwieser, 2004; Csernoch, 2014a, 2014b, 2015; Csernoch & Biró, 2015a, 2015b, 2015c, Biró & Csernoch, 2015a, 2015b). The concept of function, introduced in maths classes, and the idea that spreadsheets and spreadsheet functions are closely related to it is not mentioned at all.

We can conclude, in general, that these books are extremely contradictory; consequently, they cannot be used, either in classroom teaching, or in autonomous learning. These methods have led to risky spreadsheet documents containing formula errors routed in copying, using constants in formulas, ill-used references, incorrect selection of functions, incorrect argument list, etc. In general, the documents lack of design and concept, which unplugged phases are rarely taught either in schools or in additional materials for lifelong learners (Angeli, 213; Raffensperger, 2001; Thorne, 2005; Thorne, 2010). Further consequences of the 'classical' teaching approach and materials are the time, human and computer resources used up (Van Deursen & Van Dijk, 2012), and overconfidence (Kruger & Dunning, 1999; Thorne, 2005). Since schema construction is not preferred in these methods, applying them is extremely demanding – considering cognitive load –, with all its consequences (Thorne, 2005).

One further problem has to be mentioned here. Analyzing textbooks and publishing the results would improve their quality. However, we have experienced that some authors consider themselves so highly qualified that they refuse to accept these analyses (Csernoch 2014b; Koreczné, 2014).

Panko and Port claimed that CS should take end-user computing seriously – problem (2). We go two steps further back: (1) we claim that CS should take teaching and teacher education seriously, and (2) teaching and teaching education should also take end-user computing seriously. However, not in the way suggested by Gove (2012), by banishing end-user computing from education. If we did so, we would increase the number of self-trained end-users, who would accept the 'user-friendly' approaches of profit oriented software companies, spending lots of time learning the interfaces. We argue that teaching concept-based problem solving and schemata construction requires professionals.

## 4 ATM VS. AUM THINKING

### 4.1 Mathability levels of problem solving approaches

Panko (2013) claimed that AUM thinking would reduce the number of errors in end-user computing, while Gove (2012) and Bell & Newton (2013) asserted that routine activities kill the algorithmic aspect of these tools and activities. This contradiction tells us that teachers have to find the right proportion of ATM and AUM thinking. We have found that a typology of computer problem solving approaches (Csernoch & Biró, 2015a) and the mathability levels of software tools (Biró & Csernoch, 2015a, 2015b) would provide guidelines.

There is a close connection between the higher mathability level approaches and the types of thinking required, from Level 5 to 3. Level 5 is the concept based approach, which requires the most ATM thinking. At this level a real world problem is presented and, following Pólya's problem solving guide (1954) (see also Thorne, 2005 and Gross et al., 2014), we can reach a satisfactory result. At Level 4 the original world problem is somewhat simplified and the problem solving process starts from the building of the algorithm(s). In any other aspects there are no differences between Levels 5 and 4.

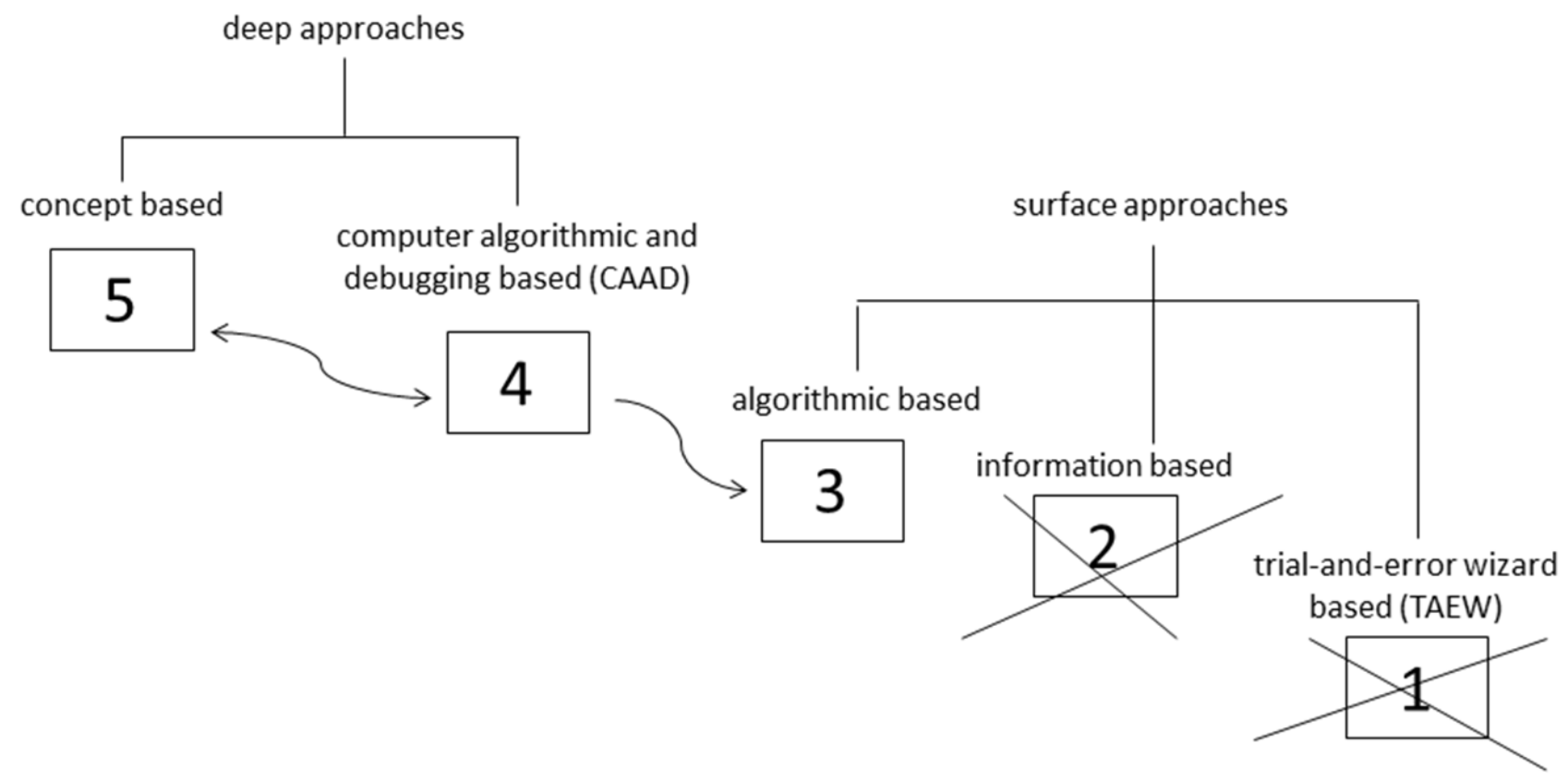

**Figure 1. Computer problem solving approaches matched with the mathability level of problem solving and software tools**

Methods at Level 2 focus on the details of the language and the environment, while at Level 1 the unplanned surface browsing leads to some kind of output. Neither of these levels is considered a problem solving approach, but rather a planned or unplanned sequence of clicks. Operating at Level 2 and/or 1 would lead to misconceptions; Sewell & Thede (2011) clearly stated that "spreadsheet languages are terse - hard to document and hard to read, hard to debug, and suitable for short subroutines or macros".

Level 3 plays a crucial role in the problem solving process, since it connects the deep and surface approach methods. The major characteristic of this level is the application of the users' own schemata – algorithms –, which is the platform where the proportion of ATM and AUM thinking can be controlled. At Level 5 and 4 ATM thinking is dominant; however, building schemata (Merribenboer & Sweller, 2005; Chi et al., 1982) at these levels would lessen the strain and the burden of ATM thinking. The schemata construction with high mathability level approaches would lead to AUM thinking, and consequently to fewer erroneous end-user activities. However, schemata construction requires teaching methods which hardly exist in end-user computing, or in spreadsheet development. This approach is well accepted in teaching 'serious' programming, but not in other computer related activities. We can also find effective schemata construction methods used in teaching maths to young children (Kemp, 1971; Morgan et al., 2014). "Routine practice is the strongest educational practice that teachers can use in their classroom to promote achievement gains," From these practices in teaching programming and maths for beginners, we can adapt these methods to educate end-users.

### 4.2 SPREGO: from ATM to AUM thinking

We claim that spreadsheet environments are as good as 'serious' programming languages, both for high mathability problem solving and for building schemata. Based on previous (Booth, 1992; Warren, 2004; Sestoft, 2011) and parallel research results (Hubwieser, 2004; Schneider, 2005) we have introduced Sprego – Spreadsheet Lego – and developed a complete methodology for introducing and teaching spreadsheets with this approach. Sprego is a Level 5 mathability approach, focusing on real world problems in various contents, which adapts the problem solving method of Pólya (1954), accepted in other sciences and also in programming, detailed in Thorne's paper (2005). The other feature of Sprego is schemata construction. Sprego introduces only a limited number of general purpose functions – a dozen for beginners – (Csernoch & Balogh, 2010; Csernoch, 2014a; Csernoch & Biró, 2015b, 2015c). Based on these functions, not only is ATM thinking and real world problem solving supported, but routine algorithms are developed, and based on them, meta-schemata are constructed. With the schemata construction ability of Sprego we can transfer knowledge from Level 4 to 3 on the mathability scale.

The limited number of Sprego functions is in accordance with findings in programming and 'classical' spreadsheets. Hromkovic claimed (2008) that "One can learn programming by starting with five instructions only and working totally with about fifteen instructions that are sufficient for programming any complex behavior of the [Logo] turtle. Our philosophy is to follow the history of programming, and so to derive all complex instructions as programs consisting of a very small set of basic instructions." Considering spreadsheet environments, Walkenbach (2010) found that "People in average do not use more than a dozen functions." With Sprego we are within the limit of 12–15 functions and have adapted the methods which have proved effective and efficient in teaching programming. We have also found that the guidelines for Logo programming would match our requirements, since, similar to Logo, the idea of Sprego is "…not to completely replace a programming course in a high-level language", but to introduce programming and algorithms and offer a tool for end-user computing. "Spreadsheets are code." (McKee; 2015) and we have to support this fact with our teaching approaches.

Conrad Wolfram, in his speech at TEDGlobal 2010 (Technology, Entertainment and Design), emphasized that in maths classes we have to "Stop Teaching Calculating, Start Teaching Math—Fundamentally Reforming the Math Curriculum". This statement is in complete accordance with the idea that "the stronger the belief in the importance of computation and correct answers the lower the mathematical content knowledge." (Francis et al., 2015). We claim that the same is true for end-user computing: Stop Teaching Software Usage, Start Teaching Computer Problem Solving—Fundamentally Reforming the Informatics Curricula.

Similar to Conrad Wolfram's ideas, Sprego is a completely new approach to teaching spreadsheet management and introductory programming in already existing environments. Sprego does not start with the introduction of the interface, the different settings of spreadsheet interfaces, saving and opening documents, or entering data. This is not spreadsheet knowledge, but general ICT skills, digital literacy, or digital competence, which are brought into Sprego classes and practiced thoroughly, but do not constitute learning objectives.

### 4.3 SPREGO: tool for functional data modelling

What Sprego stands for is in complete accordance with Hubwieser's (2004) and Schneider's (2004, 2005) theories: "Some reader may wonder why functional data modeling opens the mandatory subject informatics in the 8th grade, since until now the so called classical way was favored, i.e. the teaching of some "hard" programming skills, namely imperative-like control structures. Moreover, one will be reminded through the attribute "functional" to the paradigm of functional programming. … one has to emphasize that functional modeling is pure sequential modeling technique. Only the causal structure and the functional data flow of a context can be represented. On the other hand, a new empirical study on the learning process of students at university level has shown that students have lowest problems with the functional modeling technique but greater problems with imperative one. So it is obvious to start yet at school with functional data modeling."

Schneider (2005) did not know about the similar results which Booth obtained in 1992. However, it is heartwarming that they came to the same conclusion, unaware of each other's work. Two researchers from different surroundings, using different measuring methods achieved similar results, which supports the claim that functional programming is perfect for beginners. Our team went one step further and provided a methodology based on this theoretical background.

### 4.4 Teachers: ATM and AUM thinking

The bad news is that the proportion of ATM and AUM thinking is not constant. The proportion which would work well in end-user computing would not be appropriate in end-user teaching. Using an application and teaching it require different skills, different approaches, and different thinking modes. Teachers should be open-minded and they should be able to recognize when it is time for change. The teaching methods and the coursebooks clearly demonstrate that teachers apply AUM thinking even if it obstructs their introduction of novel teaching approaches. This problem of teachers' beliefs and their effectiveness in the teaching process is well presented in Chen et al. (2015), and the relationship between teachers' beliefs and effectiveness and students' results are presented in Figure 2 (published in the paper of Chen et al., 2015).

Based on our testing project (TAaAS, Biró & Csernoch, 2013b, 2014b; Csernoch et al., 2015; Biró et al., 2015a, 2015b) we have found that most students are prepared only for tests which rely heavily on surface-knowledge (Hromkovic, 2009), and they fail when language- and interface-independent problems are presented (Csernoch et al., 2015). Students' progress in informatics does not reach the required level of effectiveness. Beyond this, we have to keep in mind that we are not necessarily preparing students for tests, but for performing in real life, which is even more demanding than classroom and testing environments. Consequently, although the beliefs held by teachers specialized in informatics require further testing and analysis, it is clear that they have a negative impact on end-user computing.

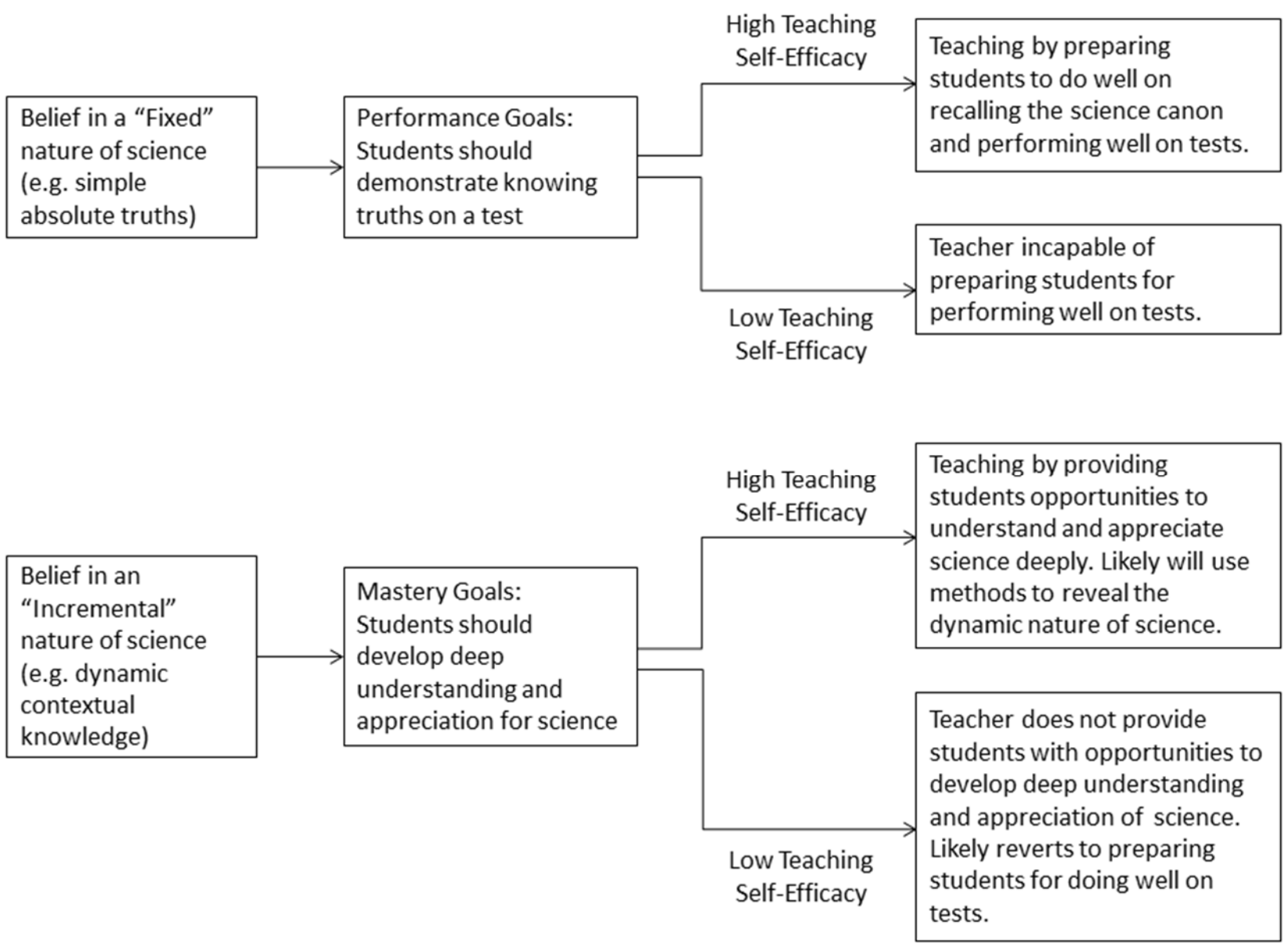

**Figure 2. The meaning system model of Chen et al. (2015). Two different beliefs in the nature of science (fixed and incremental natures, FN and IN, respectively) and two self-efficacy (high and low teaching self-efficacy, HSE and LSE, respectively) are distinguished.**

The performances of the four groups of students (Figure 2, right cells) are closely related to the mathability level of computer problem solving approaches (Figure 1): Level 3—FN+HSE, Level 2 and 1—FN+LSE, Level 5—IN+HSE, Level 4—IN+LSE, from top to bottom, respectively.

What we most miss from the teaching of end-user computing is the appreciation of this science. Panko and Port (2013) found that end-user computing is not taken seriously, "seems to be invisible…" and "It is time to stop ignoring end-user computing in general and spreadsheets in particular." We claim that the main reason for this misconception is that education is not prepared for end-user teaching. Most of our teachers do not use the algorithmic approach to end-user computing, and their teaching materials are not high mathability tools. Teachers fall for the software companies' misleading 'user-friendly' slogans and approaches, focus on technical details, develop low mathability level materials or unconditionally accept them.

## 5 CONCLUSION

Is there any reason for being optimistic or should we give up? Are end-users second hand participants in the digital word, as Asimov predicted, when he wrote that "Paul knew mysterious things about what be called electronics and theoretical mathematics and programming. Especially programming. Nicole didn't even try to understand when Paul bubbled over about it." (Asimov, 1982)? Should we also accept that "…Excel is broken. And I strongly suspect it can't be fixed. Yet it's ubiquitous and business critical. We need to reinvent the wheel and change all four whilst the car is driving down the motorway — and I don't know how to do that…" (McKee, 2015).

We cannot give up! We have to find ways for teachers to educate for effective end-user computing, especially spreadsheet management. The good news is that we already have the theoretical background (Booth, 1992; Hubwieser, 2004, Warren, 2004; Merribenboer, & Sweller, 2005) and methods (Csernoch, 2014a; Csernoch & Biró, 2015b, 2015c) which would allow us to increase the level of end-user computing, and end-users' computational thinking. The effectiveness of Sprego (Csernoch, 2014a; Csernoch & Biró, 2015b, 2015c) has been testing since the academic year of 2011/2012. The preliminary results (Biró & Csernoch, 2014) clearly demonstrate that with Sprego we can change the students' approach to spreadsheet problem solving and solutions. These results suggest that we are able to solve the problem of "changing all four whilst the car is driving down the motorway ". On the other hand, it is already clear that schemata are built with Sprego, which is necessary for the reliable decisions of fast thinking mode.

One might ask, why spreadsheets? The answer lies in their special characteristics. On the one hand, spreadsheet management is end-user activity, while on the other hand it is a form of programming. By accepting this two-fold approach in the teaching-learning process, we would raise end-users' skills and end-user computing to a higher level.

# 7 Appendices

**Appendix 1.**

**Inhalt**

**Appendix 2.**

**Excel Formulas**

A formula is a set of mathematical instructions that can be used in Excel to perform calculations. Formals are started in the formula box with an = sign.

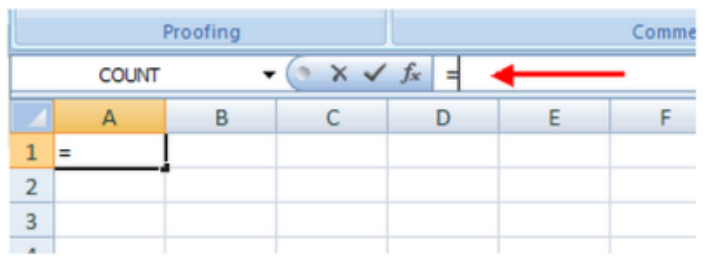

There are many elements to and excel formula.

**References:** The cell or range of cells that you want to use in your calculation
**Operators**: Symbols (+, -, *, /, etc.) that specify the calculation to be performed
**Constants**: Numbers or text values that do not change
**Functions**: Predefined formulas in Excel

To create a basic formula in Excel:

- Select the **cell** for the formula
- Type = (the equal sign) and the **formula**
- Click **Enter**

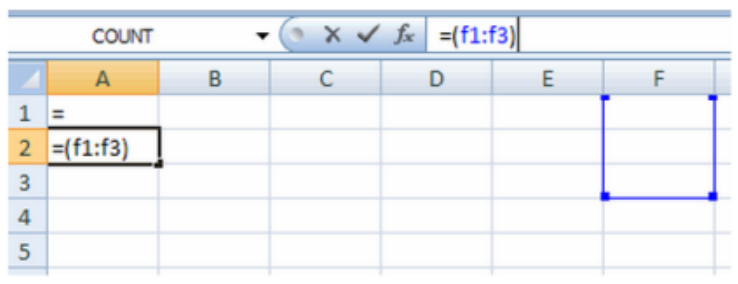

**Appendix 3.**

**Change a Cell Reference to an Absolute or Mixed Cell Reference**

1. Type the cell reference in your formula, or click the cell you want to include in the formula.
2. Type the dollar sign(s) needed to make the cell reference absolute or mixed.

*OR*

1. Type the cell reference in your formula, or click the cell you want to include in the formula.
2. Press the F4 function key to add the dollar sign(s) needed to make the cell reference absolute or mixed. The F4 function key cycles through the display of dollar signs. If you press F4 once, the cell reference changes to an absolute reference, with dollar signs before both the row and column. If you press F4 twice, only the row is made absolute, and pressing F4 a third time makes only the column absolute. Pressing F4 again clears all absolute references to the cell address.

**Copy Formulas into Noncontiguous Cells**

1. Select the cell or cells being copied.
2. To copy the cell contents to the Clipboard, click the Copy button in the Clipboard group on the HOME tab on the ribbon.
3. Click in the top, leftmost cell of the range where you want to copy the selection. Use the Paste button in the Clipboard group to paste the contents into the desired cell(s), or press the Enter key.

You can access the Copy and Paste commands from a shortcut menu by right-clicking the selection. Another method is to use the shortcut keys Ctrl+C (copy) and Ctrl+V (paste). There are shortcut keys for many of the frequently used commands. When you hover the pointer over a tool or command, its shortcut key is noted in parentheses in the ScreenTip that appears.

**Appendix 4.**

S1. sum
S2. average
S3. min
S4. max
S5. if
S6. index
S7. match
S8. vlookup
S9. hlookup
S10. count
S11. counta
S12. countif
S13. sumif
S14. averageif
S15. countifs
S16. sumifs
S17. round1
S18. small
S19. large
S20. left
S21. right
S22. len
S23. search
S24. iserror
S25. countblank
S26. value
S27. middle
S28. stdev
S29. or
S30. and
S31. dget
S32. dsum
S33. sumproduct
S34. concatenate
S35. hour
S36. perc
S37. round
S38. date
S39. text
S40. floor
S41. roundup
S42. rounddown
S43. power
S44. sqrt
S45. today
S46. lookup
S47. char
S48. replace
S49. dcount
S50. frequency
S51. year
S52. month
S53. day
S54. time
S55. product
S56. rand
S57. abs
S58. fact
S59. dmin
S60. daverage
S61. dmax
S62. median
S63. degrees
S64. isnumber
S65. log
S66. modus
S67. pmt
S68. sin
S69. rank
S70. int
S71. days360
S72. column
S73. row
S74. substitute
S75. modus
S76. var
S77. iferror
S78. now
S79. cos
S80. isna
S81. weekday
S82. cell
S83. rate
S84. second
S85. not
S86. quartile
S87. trim
S88. datevalue
S89. arcsin
S90. arccos
S91. dget
S92. trunc
S93. tan
S94. dstdev
S95. dcounta
S96. dproduct
S97. choose
S98. hyperlink
S99. critbinom
B100. pi
B101. lookup
B102. irr
B103. exact
B104. avedev
B105. dec2bin
B106. fv
B107. IGAZ
B108. HAMIS
B109. rept
B110. lower
B111. upper
B112. proper
B113. combin
B114. code
B115. offset
B116. timevalue
B117. info
B118. decimal
B119. sign
B120. linest
B121. pv
B122. ddb
B123. exp
B124. roman
B125. correl
B126. norminv
B127. ln
B128. log10
B129. permut
B130. radians
B131. subtotal
B132. sumsq
B133. geomean
B134. harmean
B135. forecast
B136. nper
B137. averagea
B138. networkdays
B139. find
B140. istext
B141. trend
B142. npv
B143. ispmt
B144. rows
B145. daysinmonth
B146. days
B147. weeksinyear
B148. accrint
B149. amordegrc
B150. arctan
B151. type
B152. averageifs
B153. duration
B154. eastersunday
B155. vdb
B156. sln
B157. db
B158. ppmt
B159. ipmt
B160. dstdevp
B161. error.type
B162. ceiling
B163. even
B164. odd
B165. transpose
B166. usdollar
B167. na
B168. effect
B169. isblank
B170. sumpozitive
B171. purecount

**Appendix 5.**

*Blank Page*